\PassOptionsToPackage{pdfpagelabels=false}{hyperref}

\documentclass{IOS-Book-Article}

\usepackage{mathptmx}
\usepackage{soul}\setuldepth{article}

\usepackage{amsmath}
\usepackage{graphicx}
\usepackage{booktabs}
\usepackage{array}
\usepackage{threeparttable}
\usepackage{tikz}
\usetikzlibrary{arrows.meta, positioning, calc, shapes.geometric}

\def\hb{\hbox to 11.5 cm{}}

\begin{document}

\pagestyle{headings}
\def\thepage{}

\begin{frontmatter}

\title{Scalable PII Discovery in Mobile App Databases via Hypothesis-Driven Search}

\author[A]{\fnms{Jeel~Piyushkumar} \snm{Khatiwala}%
\thanks{Corresponding Author: Jeel Piyushkumar Khatiwala, University of Baltimore, Baltimore, MD, USA; e-mail: Jeel.khatiwala@ubalt.edu.}},
\author[A]{\fnms{Samad} \snm{Afolabi}},
\author[B]{\fnms{Ruoyao} \snm{Xiao}},
\author[C]{\fnms{Yu} \snm{Luo}},
\author[B]{\fnms{Dianxiang} \snm{Xu}}
and
\author[A]{\fnms{Weifeng} \snm{Xu}%
\thanks{Corresponding Author: Weifeng Xu, University of Baltimore, Baltimore, MD, USA; e-mail: wxu@ubalt.edu.}}

\runningauthor{Khatiwala et al.}
\address[A]{University of Baltimore, Baltimore, MD, USA}
\address[B]{University of Missouri-Kansas City, Kansas City, MO, USA}
\address[C]{University of Central Missouri, Warrensburg, MO, USA}

\begin{abstract}
Discovering personally identifiable information (PII) in mobile forensic
databases is difficult because the relevant table-column regions are unknown,
distributed across heterogeneous SQLite schemas, and may contain values
embedded in free-text or semi-structured fields. We present a
hypothesis-driven framework that treats PII localization as bounded, adaptive
search under uncertainty. An agent ranks candidate table-column regions,
probes sampled values, and maintains a memory of prior evidence, confidence
scores, and decisions to refine subsequent hypotheses. The framework separates
lightweight PII exploration from targeted extraction, normalization, and
deduplication over validated regions, thereby limiting exhaustive inspection
to regions supported by sampled evidence. We evaluate the framework on 25
SQLite databases from 10 Android and iOS applications in the Cellebrite CTF
corpus, targeting email addresses, phone numbers, domain names, person names,
and postal addresses. Against a corpus-level distinct ground-truth set of
3,751 entities, Gemini~2.5~Pro achieves 94.5\% F1 while reducing the effective
extraction search space by 79.9\% on average. Results across 12 model backends
show strong performance among several frontier models, but substantial
sensitivity to model capability.
\end{abstract}


\end{frontmatter}

\markboth{}{}

\section{Introduction}

Mobile devices are central evidence sources in digital investigations, storing
large volumes of data across heterogeneous, app-specific SQLite databases with
frequent schema changes and limited documentation~\cite{Anglano2017,Cheng2018}.
Personally identifiable information (PII), such
as email addresses, phone numbers, domain names, person names, and postal
addresses, provides important pivots for linkage, attribution, and exposure
analysis. Locating it is difficult because the relevant table-column regions
are unknown, while PII values may occur in message bodies, serialized objects,
logs, caches, and auxiliary tables whose names and semantics can be
misleading~\cite{Quick2014,Lin2018}. Relevant values may also be embedded in
free-text or semi-structured fields.

The challenge is therefore not only extracting PII, but first determining where
it is stored. Application-specific methods depend on known schemas or artifact
locations, while broad scanning tools such as
\texttt{bulk\_extractor}~\cite{Garfinkel2013} examine large amounts of
potentially irrelevant content
and can produce substantial noise and analyst-review burden. Pattern-based
extraction also provides limited contextual interpretation for entities such as
person names and postal addresses. LLM-based methods improve semantic
interpretation~\cite{liu2025evaluating}, but typically assume that relevant
fields have already been localized. Long-context capabilities do not eliminate
this localization problem because large forensic datasets must first be parsed
and converted into model-consumable inputs, much of which may be irrelevant to
the target PII type. This motivates a search strategy that first identifies
promising regions and performs targeted extraction only after validation.

We address this localization problem as bounded, adaptive search under
uncertainty. Given a database, a target PII specification, and an exploration
budget, an agent generates hypotheses about promising table-column regions and
probes sampled values. Evidence, confidence scores, and decisions are retained
in memory and used to refine subsequent hypotheses. The framework separates
\emph{PII exploration}, which validates candidate regions through lightweight
sampling, from \emph{PII extraction}, which retrieves, normalizes, and
deduplicates entities only from validated regions. This separation limits
exhaustive inspection to regions supported by sampled evidence.

This work makes the following contributions:
\begin{itemize}
\setlength\itemsep{0em}

\item We formulate PII localization as bounded, adaptive search under
existence, location, and representation uncertainty, with memory-guided
refinement of region hypotheses.

\item We develop a two-stage framework combining agent-guided planning,
sample-aware exploration, confidence-based validation, targeted extraction,
and deterministic query mediation with provenance capture.

\item We evaluate the framework on 25 SQLite databases from 10 Android and iOS
applications across 12 language-model backends, measuring discovery
effectiveness, extraction-space reduction, and performance relative to
\texttt{bulk\_extractor} and Microsoft Presidio.
\end{itemize}


\section{Related Work}
\label{sec:related}

Mobile investigations commonly recover messages, contacts, and logs from
application-specific SQLite databases. FORC~\cite{daraghmi2023forensic}
automates Android SQLite identification, while application-specific studies
such as Telegram forensics~\cite{Anglano2017} rely on known schemas or artifact
locations. Pattern-based tools such as \texttt{bulk\_extractor}~\cite{Garfinkel2013}
scan data efficiently but produce noise and provide limited support for
semantically complex entities such as person names and postal addresses.
LLM-based PII extraction improves semantic interpretation~\cite{liu2025evaluating}
but generally assumes that relevant content has already been localized. Our
work instead addresses bounded localization of PII-bearing regions in
heterogeneous mobile databases.

Static approaches such as EviHunter~\cite{Cheng2018,Lin2018} infer evidentiary
storage from application code but are sensitive to application updates and
obfuscation. Recent forensic agents~\cite{Wickramasekara2024,Sharma2025} build
on tool-using LLM frameworks such as ReAct~\cite{yao2023react},
Toolformer~\cite{schick2023toolformer}, and AutoGen~\cite{wu2023autogen}.
Within digital forensics, LLM-identified evidence has been validated through
knowledge-graph cross-referencing with deterministic, traceable
identifiers~\cite{khatiwala2025reliability}, and agentic structural inference
over undocumented mobile databases can yield joins that execute yet diverge
from expert-verified structure~\cite{khatiwala2026structural}.
Because agentic systems remain vulnerable to hallucination and indirect prompt
injection~\cite{greshake2023indirect}, our design constrains database access
through deterministic read-only tools and structured outputs
(Section~\ref{sec:security}).

Commercial and open-source PII detection systems, including Microsoft
Presidio, AWS Comprehend PII, and Google Cloud DLP, combine regular expressions,
validation rules, and named-entity recognition. However, they operate on text
streams or preselected fields rather than planning over database schemas. Our
planning layer operates upstream and can dispatch localized column content to
either an LLM or a conventional extractor.

\section{Problem Formulation and System Architecture}
\label{sec:problem}

Let $D$ denote a mobile forensic database and $E$ a target PII type. A
\emph{region} is a table-column pair $r=(T,C)$, and $\sigma(r)$ denotes its
available schema metadata, including the table name, column name, declared
SQLite type or affinity, and other structural information. Each target type is
described by a specification $\phi(E)$ that captures its intended semantics,
representative formats, examples, and exclusion criteria.

Here, \emph{search under uncertainty} means locating PII-bearing regions with
no prior knowledge of whether the target type exists in $D$
(\emph{existence}), where it is stored (\emph{location}), or how it is
represented (\emph{representation}): a run may terminate without validating
any region, table-column regions are ranked and probed adaptively, and sampled
values are judged against the target specification, type-specific prefilters,
and semantic assessment rather than schema names or fixed patterns alone.

Given $D$, $E$, and an exploration budget $B$, the objective is to identify a
set of validated regions $R^\star$ containing instances of $E$ while performing
at most $B$ exploratory probes. Each probe examines sampled values from one
candidate region. The final discovery output is the union of entities extracted
from all validated regions:
\(\mathcal{S}(D,E)=\bigcup_{r\in R^\star}\operatorname{Extract}(r)\).

The search is guided by adaptive, agent-generated relevance hints rather than
fixed schema labels. At exploration step $t$, the planning agent uses the target
specification $\phi(E)$, region metadata $\sigma(r)$, and the current memory
state $M_t$ to derive a heuristic relevance hint
\(h_t(r,E)=\operatorname{Plan}(\phi(E),\sigma(r),M_t)\).
The hint is a search hypothesis used to rank candidate regions; it is not
evidence that region $r$ contains instances of $E$. The planner selects a
promising region $r_t$, and the exploration stage evaluates sampled values from
that region, producing observed evidence $o_t$, a confidence score
$c_t\in[0,1]$, and a decision
$d_t\in\{\textit{reject},\textit{confirm},\textit{validate}\}$.

Each exploration result is written to memory:
\(M_{t+1}=\operatorname{Update}(M_t,r_t,o_t,c_t,d_t)\).
Memory retains previously explored regions, sampled evidence, confidence
scores, validation decisions, rejected hypotheses, and the remaining budget.
The updated state allows the planning agent to revise earlier hypotheses and
derive new hints for the next probe. Region ranking therefore evolves as
evidence accumulates rather than remaining fixed from the initial schema.

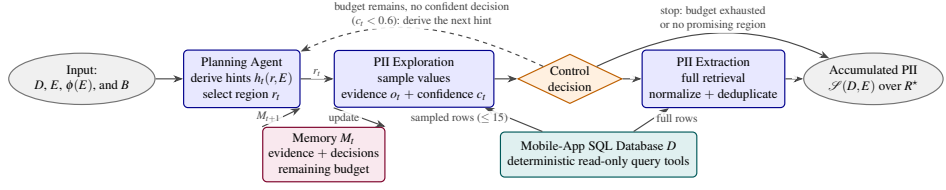
\begin{figure*}[!t]
\centering
\resizebox{\textwidth}{!}{%
\begin{tikzpicture}[
  font=\small,
  >={Stealth[length=2.4mm]},
  term/.style={
    ellipse,
    draw=black!55,
    fill=black!6,
    thick,
    align=center,
    inner sep=3pt,
    minimum height=10mm
  },
  llm/.style={
    rectangle,
    rounded corners=3pt,
    draw=blue!55!black,
    fill=blue!9,
    thick,
    align=center,
    inner sep=5pt,
    minimum height=11mm
  },
  dec/.style={
    diamond,
    aspect=2.1,
    draw=orange!75!black,
    fill=orange!12,
    thick,
    align=center,
    inner sep=1pt
  },
  tool/.style={
    rectangle,
    rounded corners=3pt,
    draw=teal!70!black,
    fill=teal!12,
    thick,
    align=center,
    inner sep=5pt,
    minimum height=11mm
  },
  mem/.style={
    rectangle,
    rounded corners=3pt,
    draw=purple!65!black,
    fill=purple!9,
    thick,
    align=center,
    inner sep=4pt,
    minimum height=10mm
  },
  flow/.style={
    -{Stealth[length=2.4mm]},
    thick,
    black!72
  },
  back/.style={
    -{Stealth[length=2.4mm]},
    thick,
    black!72,
    dashed
  },
  lbl/.style={
    font=\footnotesize,
    text=black!72,
    align=center,
    inner sep=1.5pt,
    fill=white,
    fill opacity=0.88,
    text opacity=1
  }
]

\node[term] (start) at (0,3.0)
  {Input:\\$D$, $E$, $\phi(E)$, and $B$};

\node[llm] (plan) at (3.9,3.0)
  {Planning Agent\\
   derive hints $h_t(r,E)$\\
   select region $r_t$};

\node[llm] (expl) at (8.1,3.0)
  {PII Exploration\\
   sample values\\
   evidence $o_t$ $+$ confidence $c_t$};

\node[dec] (dec) at (11.9,3.0)
  {Control\\decision};

\node[llm] (extr) at (15.4,3.0)
  {PII Extraction\\
   full retrieval\\
   normalize $+$ deduplicate};

\node[term] (done) at (19.3,3.0)
  {Accumulated PII\\
   $\mathcal{S}(D,E)$ over $R^\star$};

\node[mem] (memory) at (5.9,1.15)
  {Memory $M_t$\\
   evidence $+$ decisions\\
   remaining budget};

\node[tool] (tool) at (12.6,1.15)
  {Mobile-App SQL Database $D$\\
   deterministic read-only query tools};

\draw[flow] (start) -- (plan);

\draw[flow] (plan) --
  node[lbl,above]{$r_t$}
  (expl);

\draw[flow] (expl) -- (dec);

\draw[flow] (dec) --
  node[lbl,above,align=center]
  {}
  (extr);

\draw[flow] (extr) --
  node[lbl,above]{}
  (done);

\draw[flow] (expl.south west) --
  node[lbl,left,pos=0.55]{update}
  (memory.north east);

\draw[flow] (memory.north west) --
  node[lbl,left,pos=0.55]{$M_{t+1}$}
  (plan.south east);

\draw[back] (dec) to[out=155,in=25]
  node[lbl,above,align=center]
  {budget remains, no confident decision\\
   ($c_t<0.6$): derive the next hint}
  (plan);

\draw[flow] (dec) to[out=25,in=155]
  node[lbl,above,align=center]
  {stop: budget exhausted\\
   or no promising region}
  (done);

\draw[flow] (tool) --
  node[lbl,left,pos=0.55]
  {sampled rows ($\leq15$)}
  (expl);

\draw[flow] (tool) --
  node[lbl,right,pos=0.57]
  {full rows}
  (extr);

\end{tikzpicture}%
}

\caption{Operational workflow of adaptive, hypothesis-driven PII discovery, in which exploration results are stored in memory and used to refine subsequent region hints and search decisions.}

\label{fig:architecture}
\end{figure*}

Figure~\ref{fig:architecture} summarizes this adaptive discovery process,
showing how the planning agent generates region hypotheses from the target
specification, schema metadata, and accumulated memory; how exploration updates
the memory state; and how validated regions are passed to targeted extraction.

In our implementation, the exploration budget is $B=2$ probes per
$(\text{database},\text{PII type})$ pair, and each probe samples at most 15 rows
from one candidate region. Each probe judges whether the region contains
instances of $E$ with confidence $c_t$: a positive judgment with $c_t\geq0.6$
validates the region into $R^\star$, a negative one with $c_t\geq0.6$ rejects
it, and otherwise the planner reprobes with a revised hypothesis while budget
remains, terminating on a confident decision or an exhausted budget. Discovery
therefore proceeds in two stages:
\emph{PII exploration} adaptively evaluates candidate regions, and
\emph{PII extraction} performs targeted retrieval, normalization, and
deduplication over validated regions. 

The confidence score $c_t$ is produced by the model itself: type-specific
regular-expression prefilters restrict which values text columns contribute to
the sample, and the model returns a structured judgment over the sampled
values (found flag, confidence in $[0,1]$, reason), with unparseable
responses defaulting to zero confidence. These heuristic, uncalibrated scores,
the budget, the decision threshold, and the sampling limit were fixed
empirically to bound per-database exploration cost and held constant across
all backends.

\section{Evaluation}
\label{sec:evaluation}

All experiments instantiate the Section~\ref{sec:problem} workflow unchanged;
the research questions differ only in what is measured or, for RQ3, in the
language-model backend.

\subsection{Empirical Study Design}
\label{subsec:design}

The study uses two devices representing simulated suspects: a Samsung Galaxy
S21 running Android~14 and an iPhone~11~Pro running iOS~17.5.1, each with
256\,GB of storage and acquired through Full File System extraction. The
Android device (\emph{Otto}) contains 500 applications and 1,531 SQLite
databases; the iPhone (\emph{Sharon}) contains 158 applications and 1,504
SQLite databases. The data come from the Cellebrite Capture-the-Flag corpus
available through the NIST CFReDS
portal~\cite{pagano_evidence_locker,nist_cfrds}.

For each device, we select five applications and up to three databases per
application using file size, recency, and forensic domain knowledge, yielding
25 databases selected before content-level analysis. On Android, we include
three databases each from WhatsApp, Snapchat, Telegram, Google Maps, and
Samsung Internet (A1--A5; 15 databases). On iOS, we include three from
WhatsApp, two each from Contacts, Safari, and Calendar, and one from Apple
Messages (I1--I5; 10 databases). The resulting set spans 10 applications,
multiple application categories, and diverse schemas and storage conventions.
The evaluation therefore examines cross-schema robustness within this corpus;
it does not claim population-level representativeness of all mobile devices.

We report \emph{distinct} entities under a uniform canonicalization procedure:
email addresses are lowercased; phone numbers are digit-normalized to U.S.\
10-digit values; person names are trimmed and case-folded; and domains are
stripped of protocols and any \texttt{www.} prefix. Name matching permits
first-and-last-name inversion (\texttt{john smith} matches
\texttt{smith, john}), while postal addresses use a normalized substring rule
with safeguards against trivial matches. A strict one-to-one assignment
ensures
\(\textit{TP}\leq\min(|\mathcal{G}|,|\mathcal{S}|)\),
where $\mathcal{G}$ and $\mathcal{S}$ denote the ground-truth and system-output
sets, respectively. Results are reported at the application level unless
otherwise noted. After corpus-level canonicalization and deduplication, the
ground-truth set contains 3,751 distinct entities.

Ground truth was constructed through manual inspection of all 25 databases and
cross-checked against published Cellebrite CTF
artifacts~\cite{pagano_evidence_locker,nist_cfrds}. Under a two-stage protocol, a
practitioner generated candidates from the raw SQLite content, and a reviewer
verified the candidates and flagged normalization violations. Disputes were
adjudicated using U.S.\ phone area-code validity and domain resolvability. Of
4,766 initial candidates, 476 entries were disputed; after adjudication, 4,290
entries remained before corpus-level canonicalization and deduplication. A
third annotator then spot-checked a stratified 10\% sample across all five PII
types and found no errors in the sampled entries. The primary annotator was
blinded to system outputs and completed the annotations before any system runs.
Public figures appearing in conversational content were excluded under a rule
established before evaluation. The complete protocol is provided in the
artifact repository.

Because forensic triage places a high cost on missed relevant identifiers, we
treat recall as the primary effectiveness measure. We report precision to
capture the false-positive burden and F1 to summarize the balance between
precision and recall.

For RQ1 and RQ2, all LLM-mediated stages use
Gemini~2.5~Pro~\cite{gemini25report} at temperature 0 with fixed prompts and tool interfaces.
Type-specific regex prefilters provide surface matching for regular entity
types, while the model ranks candidate regions and interprets sampled values
during exploration. For RQ3, we hold the databases, prompts, tools, sampling
limits, confidence thresholds, stopping criteria, decoding settings, and
scoring procedure constant and vary only the language-model backend.

\subsection{RQ1: Discovery Yield, Forensic Plausibility, and Search-Space Reduction}
\label{sec:rq1}

\textbf{RQ1:} What distinct PII yield does the system achieve across diverse Android and iOS applications, are the recovered entities grounded in plausible forensic artifacts, and how much does hypothesis-driven planning reduce the extraction search space?

Table~\ref{tab:pii_per_app_gemini} reports both, per application, for Gemini-2.5-Pro. Distinct yield varies substantially, highest in communication- and identity-centric apps (iOS and Android WhatsApp, Snapchat, iOS Contacts) and smaller but non-trivial from Maps, browsers, and Calendar. Telegram (A3) yielded nothing, consistent with its encrypted storage, which exposed no interpretable plaintext. The distribution is forensically plausible: high-yield apps store dense user artifacts while lower-yield apps expose only cached metadata or auxiliary records, so the entities reflect genuine artifacts rather than incidental matches.

The same table reports search-space reduction. We distinguish the agent's
\emph{reasoning space}, comprising the full schema, from its
\emph{effective extraction space}, comprising the columns escalated to
exhaustive row-level inspection. For application $a$, with
$C_{\text{total}}(a)$ candidate columns and $C_{\text{scan}}(a)$ fully
extracted, the reduction is
\(\text{Reduction}(a)=1-C_{\text{scan}}(a)/C_{\text{total}}(a)\).
Reductions range from 64.84\% to 99.50\%, indicating that only a minority of
candidate columns are escalated. During manual inspection, we observed that PII
was often concentrated in application-specific regions associated with user
profiles, registration records, contacts, chats, and messages. This
concentration helps explain why hypothesis-driven planning can eliminate many
unrelated columns before full extraction. Messaging applications generally
permit stronger pruning, whereas Contacts, Safari, and Apple Messages require
broader inspection. The reduction metric captures column-level pruning rather
than runtime, token consumption, or computational cost.

\begin{table*}[!t]
\centering
\footnotesize
\setlength{\tabcolsep}{4pt}
\caption{Per-application results for Gemini~2.5~Pro: distinct PII discovered
by type (Email--Postal, with Total), and reduction of the effective extraction
space (total columns, columns scanned, and reduction).}
\label{tab:pii_per_app_gemini}

{\renewcommand{\arraystretch}{0.85}
\resizebox{\textwidth}{!}{%
\begin{tabular}{@{}clrrrrrrrrr@{}}
\toprule
\textbf{ID} & \textbf{Application} & \textbf{Email} & \textbf{Phone} &
\textbf{Domain} & \textbf{Person} & \textbf{Postal} & \textbf{Total} &
\textbf{Cols} & \textbf{Scan} & \textbf{Reduct.} \\
\midrule
A1 & WhatsApp & 0 & 355 & 45 & 81 & 9 & \textbf{490} & 1627 & 118 & 92.75\% \\
A2 & Snapchat & 2 & 50 & 148 & 261 & 0 & \textbf{461} & 842 & 90 & 89.31\% \\
A3 & Telegram & 0 & 0 & 0 & 0 & 0 & \textbf{0} & 1197 & 6 & 99.50\% \\
A4 & Google Maps & 0 & 5 & 6 & 4 & 8 & \textbf{23} & 71 & 14 & 80.28\% \\
A5 & Samsung Internet & 1 & 0 & 20 & 2 & 0 & \textbf{23} & 173 & 54 & 68.79\% \\
I1 & WhatsApp & 0 & 1655 & 30 & 1090 & 2 & \textbf{2777} & 328 & 75 & 77.13\% \\
I2 & Contacts & 6 & 292 & 30 & 0 & 3 & \textbf{331} & 219 & 77 & 64.84\% \\
I3 & Apple Messages & 7 & 33 & 22 & 16 & 0 & \textbf{78} & 181 & 52 & 71.27\% \\
I4 & Safari & 0 & 0 & 17 & 2 & 0 & \textbf{19} & 72 & 24 & 66.67\% \\
I5 & Calendar & 1 & 0 & 4 & 0 & 0 & \textbf{5} & 539 & 60 & 88.87\% \\
\bottomrule
\end{tabular}%
}}
\end{table*}

\subsection{RQ2: Robustness Across PII Types}
\label{sec:rq2}

\textbf{RQ2:} How robust is the proposed framework across PII types in terms
of corpus-level distinct recall and precision?

Let $\mathcal{G}$ denote the corpus-level distinct ground-truth set, containing
3,751 entities, and let $\mathcal{O}$ denote the distinct entities returned by
the proposed framework. We compute recall as
\( |\mathcal{G}\cap\mathcal{O}|/|\mathcal{G}| \) and precision as
\( |\mathcal{G}\cap\mathcal{O}|/|\mathcal{O}| \).

\begin{table*}[!t]
\centering
\footnotesize
\setlength{\tabcolsep}{5pt}
\caption{Corpus-level distinct PII performance by type. Ground-truth entities
(GT), entities returned by the proposed framework (Output), and true positives
(TP) are reported with distinct recall and precision.}
\label{tab:pii_distinct_recall}

{\renewcommand{\arraystretch}{0.85}
\begin{tabular}{@{}lrrrrr@{}}
\toprule
\textbf{PII Type} & \textbf{GT} & \textbf{Output} & \textbf{TP} &
\textbf{Recall} & \textbf{Precision} \\
\midrule
Email Address  &   15 &   14 &   13 & 86.7\% & 92.9\% \\
Phone Number   & 2085 & 2068 & 2003 & 96.1\% & 96.9\% \\
Domain Name    &  148 &  195 &  134 & 90.5\% & 68.7\% \\
Person Name    & 1491 & 1434 & 1394 & 93.5\% & 97.2\% \\
Postal Address &   12 &   22 &   11 & 91.7\% & 50.0\% \\
\bottomrule
\end{tabular}
}
\end{table*}

As shown in Table~\ref{tab:pii_distinct_recall}, the proposed framework
maintains strong recall across all five PII types, ranging from 86.7\% for
email addresses to 96.1\% for phone numbers. Phone numbers and person names,
the two largest ground-truth categories, achieve both high recall and high
precision, indicating that the framework is effective for both structured and
context-dependent entities.

Precision varies more substantially across types. Domain-name precision falls
to 68.7\%, primarily because of boundary and canonicalization differences such
as \texttt{s.whatsapp.net} versus \texttt{whatsapp.net}. Postal addresses
achieve 91.7\% recall but only 50.0\% precision; however, the ground truth
contains only 12 postal addresses, so a small number of false positives has a
large effect on the percentage. Email performance should be interpreted
similarly because the corpus contains only 15 distinct email addresses.
Overall, the proposed framework provides robust recall across PII types, while
precision is more sensitive to entity-boundary ambiguity and small
ground-truth counts.

\subsection{RQ3: Sensitivity to Language Model Choice}
\label{sec:rq3}

\textbf{RQ3:} How does the choice of language model affect distinct discovery robustness and search-space reduction under the same workflow?

Table~\ref{tab:model_yield} compares 12 LLM backends from the GPT, Gemini,
Claude Opus, Qwen2.5, Mistral/Mixtral, and LLaMA families. Each backend runs
the same pipeline with identical databases, prompts, stopping criteria, tools,
and temperature settings and is scored against the same 3,751-entity
ground-truth set. The benchmark-score (BM) column reports published MMLU
results where available as a coarse capability proxy. We also include two
non-LLM baselines: \texttt{bulk\_extractor}~v1.6.0~\cite{Garfinkel2013},
which supports email, phone, and domain extraction, and
Presidio, a rule- and NLP-based toolkit~\cite{presidio}. These systems serve as
end-to-end references for unguided extraction rather than as competing
planners. Because Presidio could also operate downstream of localization, the
comparison provides a reference for assessing hypothesis-driven localization
and validation relative to applying a fixed extractor broadly across the
content.

\begin{table*}[!t]
\centering
\footnotesize
\setlength{\tabcolsep}{2.5pt}
\caption{Sensitivity to language model choice: distinct PII yield, discovery
effectiveness, and search-space reduction. Agentic models are sorted by F1.}
\label{tab:model_yield}
{\renewcommand{\arraystretch}{0.85}%
\resizebox{\textwidth}{!}{%
\begin{threeparttable}
\begin{tabular}{@{}lrrrrrrrrrrr@{}}
\toprule
\textbf{Method/LLM} & \textbf{BM} & \textbf{Email} & \textbf{Phone} &
\textbf{Dom.} & \textbf{Per.} & \textbf{Post.} & \textbf{Total} &
\textbf{Prec.} & \textbf{Rec.} & \textbf{F1} & \textbf{Red.} \\
\midrule
Gemini-2.5-Pro    & 89.2\% & 14 & 2068 & 195 & 1434 & 22 & 3733 & 94.70\% & 94.24\% & 94.47\% & 79.9\% \\
GPT-4.1           & 90.2\% & 15 & 2015 & 123 & 1116 &  9 & 3278 & 98.78\% & 86.32\% & 92.13\% & 85.5\% \\
Claude Opus 4.6   & N/A$^{\circ}$ & 16 & 2106 & 143 & 1674 & 8 & 3947 & 87.99\% & 92.59\% & 90.23\% & 82.9\% \\
GPT-4o-mini       & 82.0\% & 10 & 1336 &  97 & 1320 &  8 & 2771 & 97.91\% & 72.33\% & 83.20\% & 92.7\% \\
GPT-5.1           & N/A$^{\star}$ & 16 & 1146 & 137 & 1396 & 10 & 2705 & 97.86\% & 70.57\% & 82.00\% & 80.8\% \\
Qwen2.5-72B       & 86.1\% & 16 &  391 &  63 &  872 &  6 & 1348 & 96.44\% & 34.66\% & 50.99\% & 80.7\% \\
Mistral-Large     & 84.0\% &  5 &   53 & 102 &  292 &  5 &  457 & 86.43\% & 10.53\% & 18.77\% & 92.3\% \\
GPT-3.5-Turbo     & 70.0\% &  9 &   41 &  18 &  293 &  7 &  368 & 91.03\% &  8.93\% & 16.27\% & 90.9\% \\
Mixtral-8x22B     & 77.8\% & 15 &   51 &  95 &  199 &  5 &  365 & 81.37\% &  7.92\% & 14.43\% & 87.8\% \\
Mixtral-8x7B      & 70.6\% &  2 &   32 &  18 &   10 &  0 &   62 & 70.97\% &  1.17\% &  2.31\% & 92.7\% \\
LLaMA-3.1-70B-Ins & 83.6\% &  0 &    3 &  22 &    3 &  1 &   29 & 89.66\% &  0.69\% &  1.38\% & 91.0\% \\
LLaMA-3.1-8B-Ins  & 73.0\% &  2 &    0 &   0 &   15 &  1 &   18 & 94.44\% &  0.45\% &  0.90\% & 97.0\% \\
\midrule
\texttt{bulk\_extractor} & N/A & 6046 & 1035 & 159 & -- & -- & 7240 & 14.72\% & 24.83\% & 18.47\%$^{\ddagger}$ & N/A \\
Presidio          & N/A & 499 & 2377 & 2917 & 2963 & 126 & 8882 & 19.00\% & 45.03\% & 26.73\%$^{\dagger}$ & N/A \\
\bottomrule
\end{tabular}
\begin{tablenotes}[para,flushleft]
\footnotesize
\item \textbf{BM}: MMLU from official provider reports
(Gemini-2.5-Pro~\cite{gemini25report},
GPT-4.1~\cite{openai_gpt41_2025},
GPT-4o-mini~\cite{openai2024gpt4omini},
GPT-3.5-Turbo~\cite{openai2023gpt4},
Qwen2.5-72B~\cite{qwen2024qwen25},
Mixtral-8x7B~\cite{mixtral8x7b},
Mixtral-8x22B~\cite{mixtral8x22b},
Mistral-Large~\cite{mistralLarge2},
LLaMA-3.1~\cite{llama3herd2024}).
$^{\star}$No verified primary MMLU value for GPT-5.1 at time of evaluation.
$^{\circ}$No verified primary MMLU source for Claude Opus 4.6.
$^{\ddagger}$\texttt{bulk\_extractor} does not support person-name or
postal-address extraction. $^{\dagger}$Presidio v2.x with spaCy
\texttt{en\_core\_web\_lg}, threshold 0.4. Per-type counts for all backends are
released in the artifact repository.
\end{tablenotes}
\end{threeparttable}%
}}
\end{table*}

Effectiveness is strongly shaped by model choice. Gemini-2.5-Pro leads with
94.47\% F1, followed by GPT-4.1 with 92.13\% F1 and the highest precision
of 98.78\%, and Claude Opus~4.6 with 90.23\% F1 and the largest yield of
3,947 entities. GPT-4o-mini and GPT-5.1 achieve approximately 82--83\% F1
but have lower recall, while the remaining models fall below 51\% F1, reaching
0.90\% for LLaMA-3.1-8B-Ins. BM is an imperfect predictor: stronger models
generally rank near the top, but effectiveness also depends on generating useful
region hypotheses, interpreting samples, and deciding when to probe or extract.
Search-space reduction must therefore be interpreted jointly with recall.
Gemini-2.5-Pro combines 79.9\% reduction with 94.24\% recall, and GPT-4.1
combines 85.5\% reduction with 86.32\% recall. In contrast, weak models may
report high reduction because they under-explore; LLaMA-3.1-8B-Ins reaches
97.0\% reduction but only 0.45\% recall. The non-LLM baselines return much
larger volumes, 7,240 and 8,882 entities, but with low precision, 14.72\% and
19.00\%, respectively. Thus, high yield or reduction alone does not indicate
effective discovery. The workflow is backend-flexible but not
backend-agnostic.

\section{Discussion and Conclusion}
\label{sec:discussion}
\label{sec:security}

The framework narrows PII extraction to regions supported by sampled evidence.
Its reported reduction measures column-level pruning, not runtime, token cost,
API cost, or analyst effort. Structured outputs, read-only query execution, and
type-specific prefilters reduce risks from misleading schemas and adversarial
content. In a limited test with five crafted iOS Messages records, no
attacker-supplied identifiers were returned, although this does not establish
robustness against adaptive attacks.

Budgeted exploration may miss low-salience regions, and performance depends on
confidence thresholds and model capability. The current implementation has
limited support for encoded formats, deleted or cloud-backed data, non-U.S.\
phone formats, cross-application linking, and unevaluated PII types. Because
the evaluation uses 25 databases from a controlled Cellebrite CTF corpus and
one run per backend, the findings do not establish population-level
generalizability or repeated-run stability.

Overall, sampled validation substantially reduces the extraction search space
while maintaining strong discovery effectiveness. Future work will broaden
format and PII coverage and evaluate stability, cost, latency, and analyst
effort.
Discovered identifiers can also seed enrichment against open-source
intelligence sources, whose defensive effectiveness is itself an active
measurement question~\cite{emeksiz2026osint}.
Recovered entities require analyst verification before investigative or legal
use.

\paragraph{Ethics and availability.}
The study uses synthetic Cellebrite CTF data from NIST CFReDS. Code, ground
truth, prompts, configurations, logs, and evaluation scripts are available at
\url{https://anonymous.4open.science/r/IDFC-2026-Agentic_PII_Extraction-86C4/};
the source data is available through NIST CFReDS.

\textbf{Acknowledgments} This material is based upon work supported by the National Science Foundation under Grant No.~2039289.

\makeatletter
\let\@oldthebibliography\thebibliography
\renewcommand{\thebibliography}[1]{\@oldthebibliography{#1}\setlength{\itemsep}{0pt}\setlength{\parskip}{0pt}\footnotesize}
\makeatother
\bibliographystyle{vancouver}
\bibliography{sample}

\end{document}